\documentclass[a4paper,fleqn]{cas-sc}

\usepackage[numbers]{natbib}
\usepackage{xcolor}
\usepackage{ulem}

\def\tsc#1{\csdef{#1}{\textsc{\lowercase{#1}}\xspace}}
\tsc{WGM}
\tsc{QE}
\begin{document}
\let\WriteBookmarks\relax
\def\floatpagepagefraction{1}
\def\textpagefraction{.001}

\shorttitle{Renewable Energy at the South Pole}    

\shortauthors{Babinec, et al.}  

\title [mode = title]{Feasibility of renewable energy for power generation at the South Pole}

\author[1]{Susan Babinec}

\affiliation[1]{organization={Argonne National Laboratory},
            addressline={9700 S. Cass Avenue}, 
            city={Lemont},
            postcode={60439}, 
            state={IL},
            country={U.S.A.}}

\author[2]{Ian Baring-Gould}
\affiliation[2]{organization={National Renewable Energy Laboratory},
            addressline={15013 Denver West Parkway}, 
           city={Golden},
            postcode={80401}, 
            state={CO},
            country={U.S.A.}}

\author[1]{Amy N. Bender}[orcid=0000-0001-5868-0748]
\cormark[1]

% Email id of the second author
\ead{abender@anl.gov}

\author[2]{Nate Blair}
\author[2]{Xiangkun Li}
\author[1]{Ralph T. Muehleisen}
\author[2]{Dan Olis}
\author[2]{Silvana Ovaitt}

\cortext[1]{Corresponding author}

\begin{abstract}
Transitioning from fossil-fuel power generation to renewable
energy generation and energy storage in remote locations has the
potential to reduce both carbon emissions and cost.  We present a
techno-economic analysis for implementation of a hybrid renewable
energy system at the South Pole in Antarctica, which currently hosts
several high-energy physics experiments with nontrivial power needs. A
  tailored model for the use of solar photovoltaics,
wind turbine generators, lithium-ion energy storage, and long-duration
energy storage at this site is explored in different combinations with and without
traditional diesel energy generation.  We find that the least-cost
system includes all three energy generation sources and
lithium-ion energy storage.   For an example
steady-state load of 170 kW, this hybrid system reduces diesel consumption by 95\% compared to
an all-diesel configuration.  Over the course of a 15-year analysis period the reduced diesel usage leads to a net
 savings of \$57M, with a time to payback of approximately two years.  All the scenarios modeled show that the transition to renewables is highly cost effective under the unique economics and constraints of this extremely remote site.  
\end{abstract}

% \begin{highlights}
% \item Renewable energy generation at the South Pole, Antarctica is explored
% \item South Pole conditions require unique renewable technical design
% \item South Pole renewable system possible with mature, commercially-available technology
% \item Least-cost hybrid renewable system reduces annual diesel consumption and CO$_2$ by 95\%
% \item South Pole renewable energy creates positive economic impact across many scenarios
% \end{highlights}

% Keywords

%%%word count 
% Word Count: 6649\\

% %Abbreviations \& Units
% \textbf{Abbreviations \& Units}
% \begin{description}
% \item[API] application programming interface
% \item[BAU] business as usual
% \item[BESS] Battery energy storage system
% \item[CO$_2$] carbon dioxide
% \item[cm] centimeters
% \item[DOE] Department of Energy
% \item[ES] energy storage 
% \item[EV] electric vehicle
% \item[GHI] global horizontal irradiance
% \item[kg] kilogram
% \item[kW] kilowatt
% \item[kW-AC] kilowatt alternating current
% \item[kW-DC] kilowatt direct current
% \item[kWh] kilowatt hour
% \item[kWh/gallon] kilowatt hour per gallon
% \item[LCOE] levelized cost of energy
% \item[LDES] long-duration energy storage
% \item[Li-ion] lithum ion
% \item[NOAA] National Oceanic and Atmospheric Administration
% \item[NPS] Northern Power Systems
% \item[NREL] National Renewable Energy Laboratory
% \item[PV] photovoltaic
% \item[REopt] Renewable Energy Integration and Optimization 
% \item[RTE] round trip efficiency
% \item[Wh/kg] watt hours per kilogram
% \item[WTG] wind turbine generators
% \item[$^{\circ}$C] degrees Celsius
% \item[\$] dollars
% \item[\$/kW] dollars per killowatt
% \item[\$M] millions of dollars
% \end{description}

\maketitle

% Main text
\section{Introduction}\label{sec:intro}

Renewable {energy sources coupled with energy storage are enabling a
  global transition away from fossil fuel generators, reducing the associated
greenhouse gas emissions.  As renewable technologies have matured over the past
decade, the energy generated has become increasingly reliable and
affordable.   Adoption of renewable energy has
accelerated with decarbonization initiatives in many countries, for
example over 90\% Clean electricity by 2035 in the US, or even ‘Carbon
Net Zero’ and ‘Net Zero World Initiative’ at the global level \cite{solarfutures, netzero, kintnermeyer2010}.   
 Using locally-available renewable energy sources in remote
regions can be particularly impactful as transportation of fuel to these
locations can be both complex and costly, and a transition to
renewables can reduce these costs along with the greenhouse gas emissions. 

Antarctic research stations are some of the most remote
facilities on the planet, relying primarily on fossil fuel to generate
power with high reliability.  In the case of the South Pole, the
supply of fossil fuel is particularly expensive due to the complicated
transportation logistics required for its delivery.  A transition to
energy technology that uses the local solar and wind resources has the
potential to reduce both the negative economic and environmental impacts.
The Protocol on Environmental Protection in the Antarctic Treaty
specifically notes that “The protection of the Antarctic
environment…shall be fundamental considerations in the planning and
conduct of all activities in the Antarctic Treaty area” \cite{treaty}.  However,
the extreme environment and logistical constraints of Antarctica pose
singular challenges in the implementation and operation of renewable
technology.  Two unique obstacles for solar installations are the
resource availability and snow accumulation. For wind turbines,
challenges center around the extreme range of weather conditions and
the associated mechanical stresses. Some progress towards
decarbonization of the Antarctic has been made with multiple stations
incorporating renewable sources to supply a fraction of their energy \cite{lucci2022,comnap}.   If the technical challenges can be resolved, then significant
opportunities exist for further adoption of renewable sources.

Renewable energy hybrid systems in Antarctica are tailored to the
specific characteristics of each site because key factors such as
terrain and weather vary widely across the continent.  For example,
Belgium’s Princess Elisabeth Station employs both wind turbines and
solar panels to generate a 100\% renewable energy supply (132 kW).
Customizations include specially designed rotors and blades to
withstand the site’s wide range of wind speeds \cite{lucci2022,princesselisabeth}. New Zealand’s Scott Base and the
United States’ McMurdo Station share the Ross Island Wind Farm, a 990 kW installation of three wind turbines supplying tens of percent of
the total energy used by the two stations, while the Black Island
Communications Station has used wind, solar and diesel engines to
provide energy \cite{brpr2012}. The Australian Casey Station has installed
30 kW of solar panels that supply $\sim$10\% of the annual energy
consumption \cite{caseystation}.  Several groups have also performed
renewable energy feasibility studies for other stations across the
continent \cite{dechristo2016, bockelmann2022, olivier2008, boccaletti2014}.  

 The Amundsen Scott South Pole Station is a United States research
 station that operates year-round to support on-going scientific
 experiments covering a wide range of disciplines from astrophysics
 and geophysics to atmospheric and climate sciences.  Located at the
 geographic South Pole, this station experiences extreme temperatures
 with a lowest recorded temperature of -82$^{\circ}$ C.  Electrical
 power for both the station and experiments is currently supplied
 entirely by diesel generators. 

Evaluations of the technical and economic feasibility of both
photovoltaic (PV) panels \cite{williams2000} and wind turbines \cite{baringould2005B} have been previously conducted for the South
Pole.  PV panels were mounted on a building for over 400 days
(including both
austral summer and winter) and the output power monitored.  No
structural degradation was observed at the end of this period and the
output power depended on both the angle of the sun and the overall
visibility due to cloud cover or blowing snow.  For wind generation, a
previous economic analysis indicated the possibility of significant
savings compared to diesel fuel despite the relatively low wind speeds
at the South Pole \cite{baringould2005B}.  Both of these studies
support the potential of renewable energy generation at this extremely
remote site.  The cost of renewable energy generation and storage have
dramatically decreased while reliability has improved since these
studies were performed \cite{way2022}. Figure \ref{fig1} shows the reduction in
cost for different renewable technologies over the past decade \cite{windcost,windmarket2021,
  solarpvbenchmark2020,bnef}. The economic case for using renewables and storage at the South Pole is therefore even stronger now. 

% Figure
\begin{figure}[ht]
  \centering
  \includegraphics[width=5in]{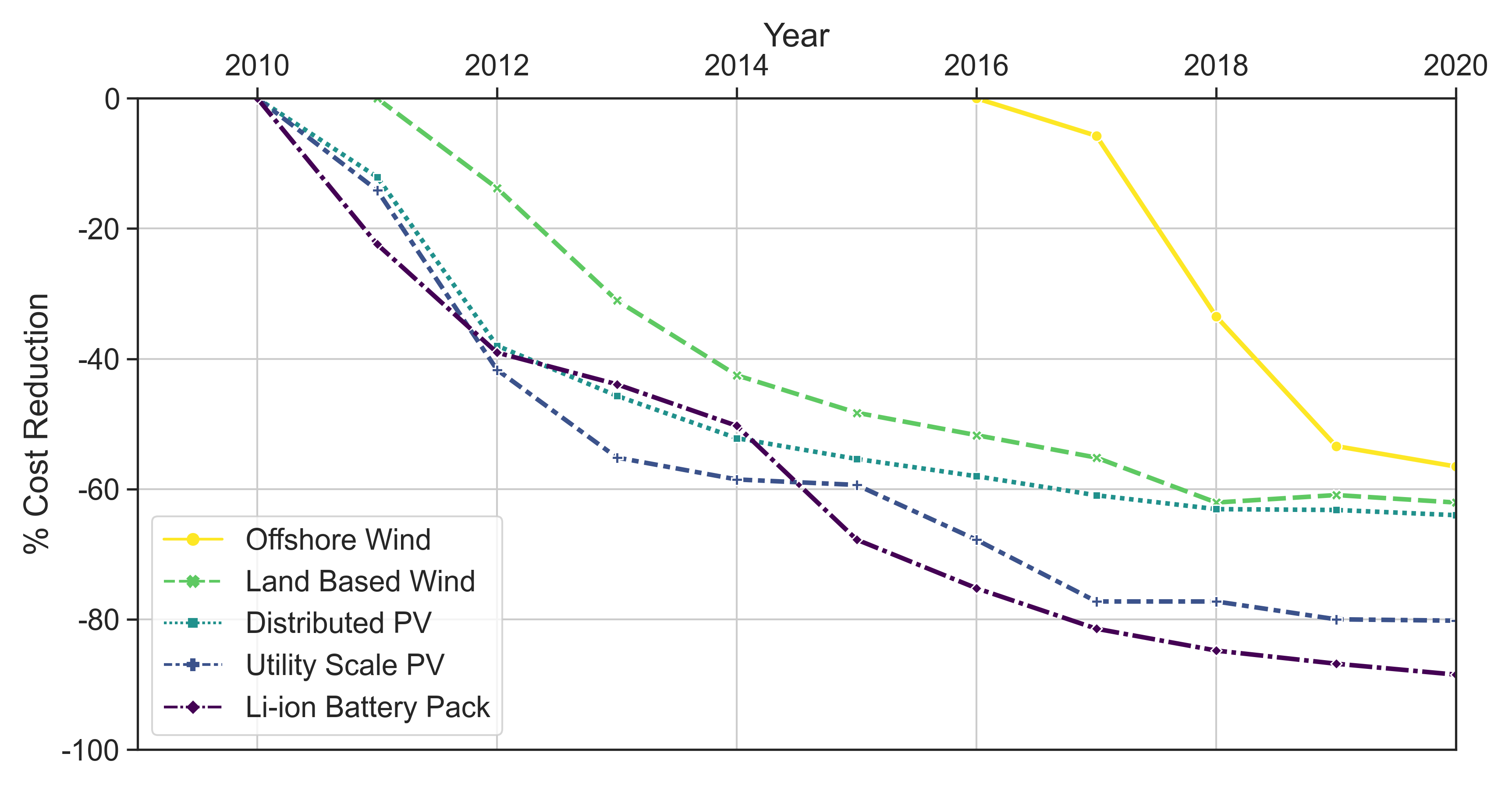}
  \caption{Summary of the cost reduction in renewable technologies
    over the past decade \cite{windcost,windmarket2021,
  solarpvbenchmark2020,bnef}.}
\label{fig1}
\end{figure}

In this work, we present an updated feasibility analysis for a South
Pole hybrid renewable energy system using solar and wind electricity
generation in combination with energy storage.  A cost optimization of
power generation technology, including existing fossil fuel
generators, is performed.   Section \ref{sec:model} introduces the renewable
technology and details how weather data measured at the South Pole
supports modeling of both solar- and wind-generated
electricity. Energy storage is also introduced, including both mature lithium-ion
batteries as well as emerging technologies for longer
durations not economically addressed with lithium-ion. A detailed cost analysis of all technologies is presented
in Section \ref{sec:reopt}, including South Pole specific costs such as shipping and
labor.  The resource and cost information are used to calculate
optimal system configurations under several different scenarios.
Discussion of the optimization results is presented  in Section \ref{sec:results}, including analysis of the
sensitivity to input assumptions.  Section \ref{sec:conclusions}
summarizes the findings and discusses a path for technical development of
the renewable-based hybrid system.

\section{Modeling Renewable Resources for the South
  Pole}\label{sec:model}

The implementation of renewable energy technology at the South Pole
site must account for the unique environmental conditions as well as
constraints imposed by the suite of scientific experiments.  The most
fundamental requirement is the ability of the equipment to survive the
extreme South Pole winter, regardless of whether it is in an
operational or standby state.  The average annual temperature is
approximately -50$^{\circ}$ C with a record low temperature is
-82$^{\circ}$ C, well below the recommended temperature ranges of most
commercial off-the-shelf renewable generation equipment.   Additional
environmental conditions include nearly zero relative humidity, an
annual snow accumulation of roughly 20 cm \cite{mosley1995}, and the
thick Antarctic ice sheet that results in a lack of solid ground.
These environmental constraints are important aspects of the solar and
wind technology identification and resource modeling discussed in the
following section.

\subsection{Solar Resource Modeling}

Solar radiation is available for energy generation at the South Pole
for six contiguous months of the year.  During that period the sun
reaches a maximum elevation of 23.5 degrees above the horizon.  The
remaining six months of the year the sun is below the horizon. Solar and other weather
data are collected by the on-site National Oceanic and Atmospheric
Administration (NOAA) South Pole
Observatory \cite{cmdlsummary} as well as by National Aeronautics and Space Administration (NASA)
satellites.  The NASA data is
made available through the POWER project\footnote{https://power.larc.nasa.gov/} with a time resolution of one
hour that is sufficient for this analysis. 
Available measurements used to model the solar resource include direct
normal irradiance, diffuse horizontal irradiance, global
horizontal irradiance (GHI), relative humidity, windspeed, and upwelling irradiance.
The ground solar radiation reflectivity, also know as albedo, can be estimated using
the upwelling irradiance with the GHI.}  The year-round permafrost contributes to high reflectivity with albedo
ranging from 0.83 to 0.97 depending on cloudiness and sun
angle \cite{Carroll1981}. The high albedo makes the South Pole an
ideal location for bifacial photovoltaic (PV) technology, which capture
irradiance from both sides of the panel.

\begin{figure}[htb]
	\centering
		\includegraphics[width=5in]{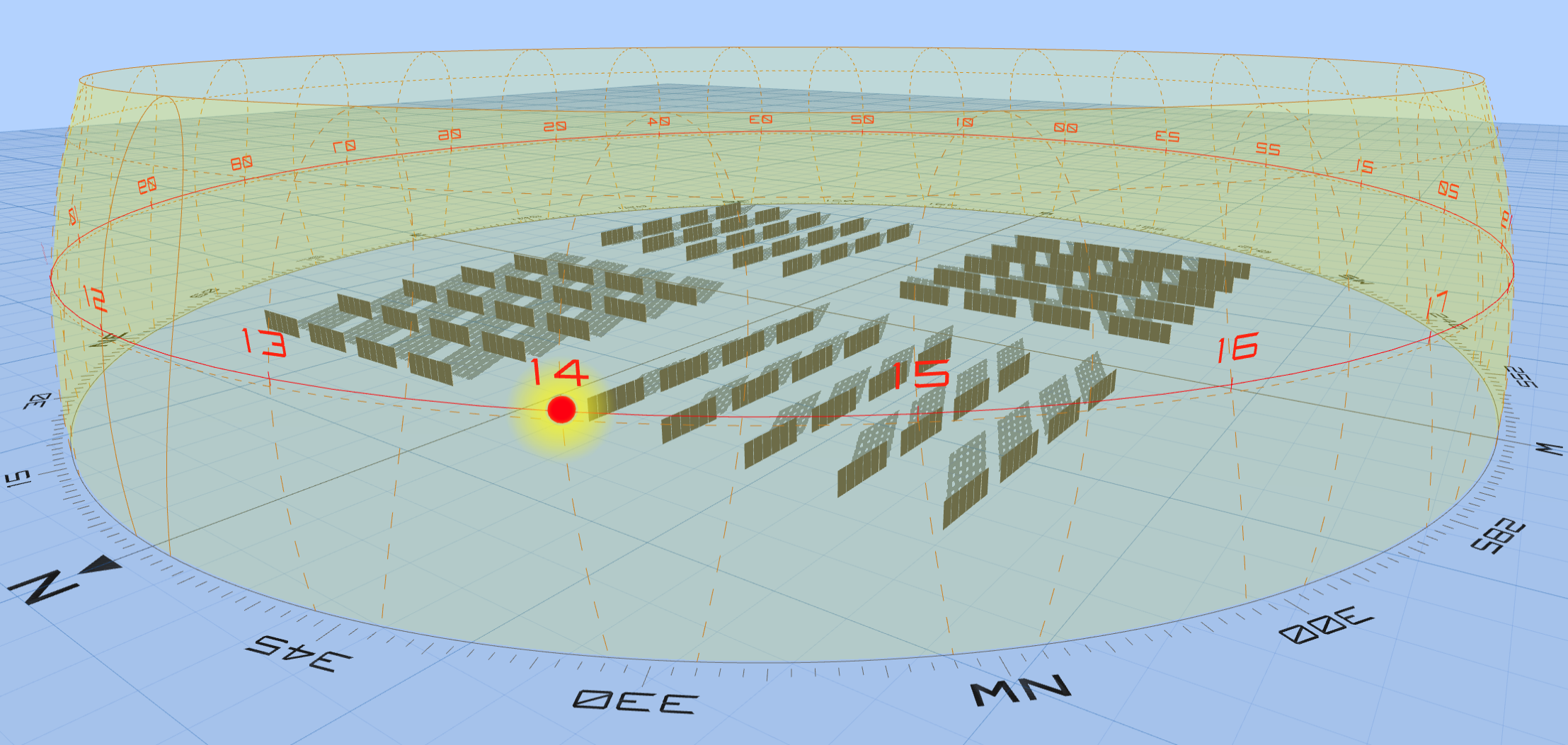}
	  \caption{Diagram of a four-directional bifacial PV 
            array. The design includes bays of six vertical mounted PV
            modules, with five bays per row separated by two meters,
            and eight rows per section.  The array depicted
            represents an 180 kW system, which is the optimized size
            for the least-cost hybrid system discussed in subsequent
            sections.  Shading shown corresponds to February 20th, 2 PM GMT-12:00. Sun position, along with current day path line, annual hourly analemma, and solar path area are also depicted.}\label{fig2}
\end{figure}

\begin{figure}[htb]
	\centering
		\includegraphics[width=0.9\textwidth]{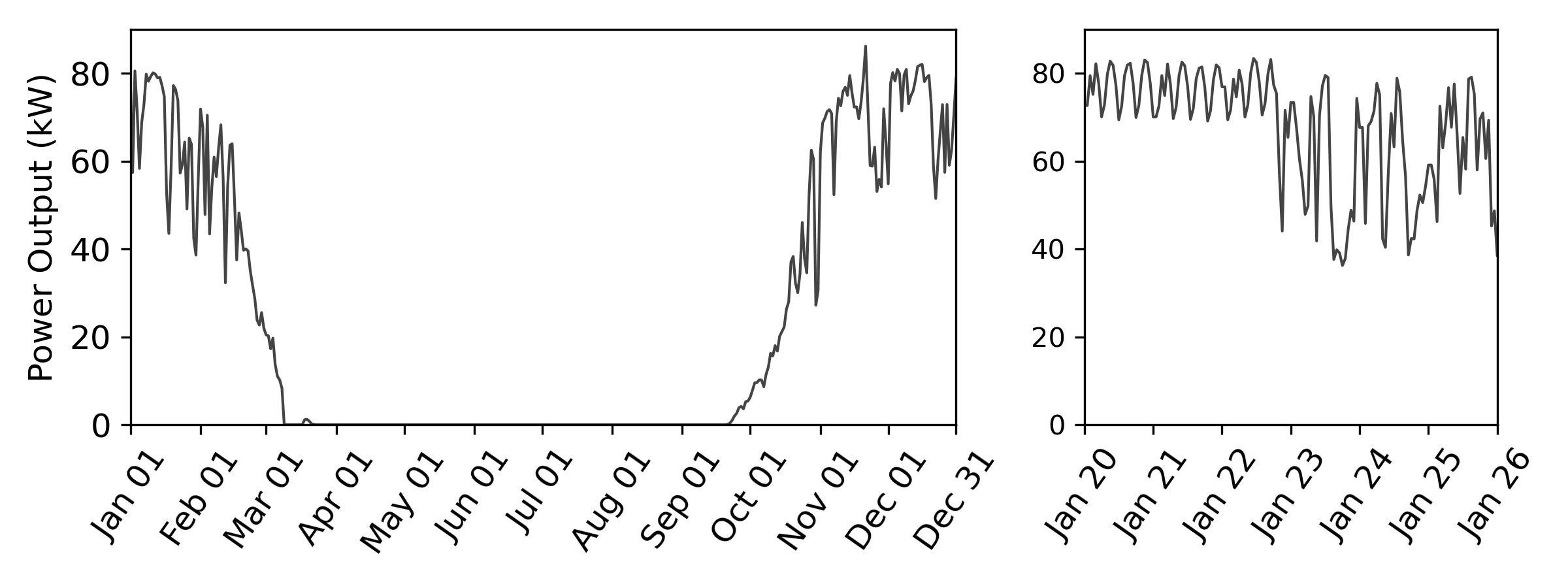}
	  \caption{(Left) Daily average of estimated energy production of 
            PV modules in the vertical array configuration for an
            example size of 100 kW-DC over one year.  The
          model uses weather data from 2016. (Right)
          Hourly energy production profile for six days;
          showing variability due to both the intrinsic array
          configuration and clouds.}\label{fig3}
\end{figure}

Solar irradiance for an array of vertical, bifacial PV panels is
modeled using the NREL open-source \textit{bifacial\_radiance} ray
tracing tool.  This ray tracing tool can
analyze complex scenarios, including tracker-element shading \cite{AyalaPelaez2020}.  A unique solar array is designed to adapt to the
  unconventional solar availability at the South Pole.  To capture the solar radiation
throughout each 24-hour revolution of the sun around the horizon the panels are
arranged into four subarrays oriented in a North-South-East-West
configuration as shown in Figure \ref{fig2}.
Modules are grouped into bays within each row at a height of 0.6m
above grade and a vertical orientation to minimize snow accumulation.
NOAA data from 2016, an average year in terms of weather, is used
to calculate the irradiance on an hourly cadence.  The simulated
irradiance is then combined with PV panel electrical parameters and the
energy generation performance of the array is determined using the
{PVlib} implementation of the single-diode model \cite{Holmgren2018}.
A Longi 72HBD-380M monocrystalline bifacial module with 0.7
bifaciality was selected as an example module owing to its good
reliability under NREL studies \cite{bifiPV2020} and placement as a
top performer on the PV Evolution Labs PV Module Reliability
Scorecard\footnote{https://modulescorecard.pvel.com/}.   Figure
\ref{fig3} shows the total output of an example 100 kW-DC solar array as a function of time.

\begin{figure}[htb]
	\centering
		\includegraphics[width=5in]{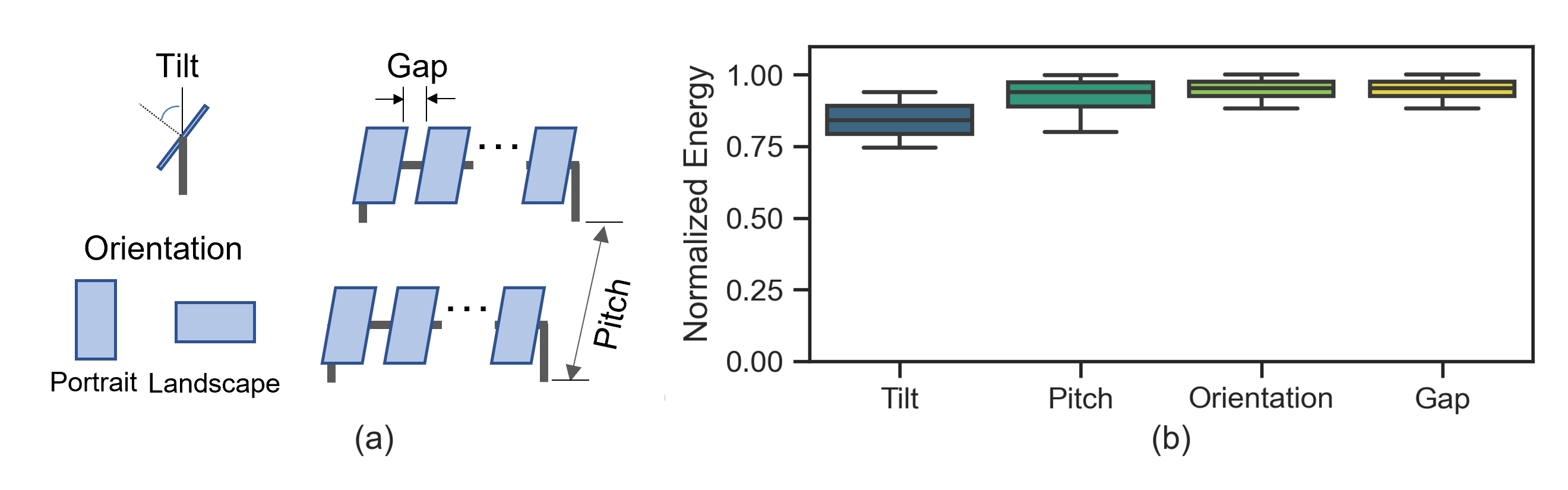}
	  \caption{(a) Conceptual diagram of the array parameters explored. (b)
            Comparison of the normalized yearly power
            production performance between variations in array
            parameters.  The baseline scenario carried forward
            includes portrait-oriented vertical panels with a
            0.6-meter ground clearance.  Panels are grouped into bays
            of six modules with 8-meter pitch between bays.  All
            configurations show similar results, with the exception of
            tilt angle where energy production decreases slightly.}\label{fig4}
\end{figure}

Several variations on the array configuration are modeled to explore
the sensitivity to the specific design. 
Initially, an array of
horizontally oriented panels (i.e., parallel to the ground) was
considered. Solar output is more constant in this
case than the vertical orthogonal array. However, the low solar angle
relative to the panel face normal resulted in significantly lower
output per unit area. Coupled with the fact that a horizontal
orientation would be more prone to snow accumulation, this
configuration was quickly eliminated from consideration.
For the vertical array configuration, variations on module
orientation, spacing between bays  of modules, and spacing between
rows in the array are explored.  An array tilted by 30
degrees from vertical is also simulated for comparison. Figure \ref{fig4} shows the resulting total annual irradiance for each
variation.  As expected, increasing the separation between rows
reduces self-shading and slightly increases cumulative irradiance.
The difference in energy production between these different
configurations is marginal, with the exception of the tilted panels,
which has significantly reduced cumulative irradiance.   The 
configuration with vertical, portrait-oriented panels is selected as
the baseline for the remainder of this study. Each bay groups together
six modules with rows that are separated with eight meter pitch.  The energy
production as a function of time for this configuration serves as an
input to the co-optimization described in Section \ref{sec:reopt}.

% Figure

\subsection{Wind Resource Modeling}
\label{subsec:wind}

Wind energy has been used extensively in isolated arctic environments,
typically operating in conjunction with diesel driven generators \cite{arcticenergy, holdmann2022, dewitt2021, remotewind}. Wind
turbines are increasingly used in these locations due to the high
energy density as compared to solar energy and the ability to provide
year-round power generation. As previously stated, wind turbines have
been used extensively in Antarctica, but only small turbines have been
installed at the South Pole, primarily to power smaller research
projects or as part of short-term demonstration projects. For power
systems at the scale of 100 kW or more, larger turbines are
required.   In this analysis, we use a 100 kW turbine which balances
turbine size, for transportation and logistics reasons, and reasonable
power output.

Wind speed data is collected by the NOAA South Pole Observatory;
however, this data has a measurement height of 10 meters above ground
level and extrapolation must be performed to determine the potential
for wind energy.   For this analysis, wind speed at the 10-meter
height along with the temperature and pressure measurements from 2003
are used at a measurement interval of one minute
\cite{baringould2005B, riihimaki2023baom}.  Separately, average
monthly reporting of wind speed data from February 2008 to June of
2009 from the NOAA meteorology tower at a height of 30 meters is used
to assess wind shear, which describes how the wind speed changes with
 distance away from the earth’s surface.  Using the
 Windographer\footnote{https://www.ul.com/software/windographer-wind-data-analytics-and-visualization-solution}
 software program, the one-minute data is converted into an estimated
 hourly average wind turbine power output for a Northern Power System
 NPS 100C-24\footnote{https://www.nps100.com/wp/nps-100c/}.  This 95 kW
 Arctic wind turbine has a 24.4-meter rotor diameter and a 30-meter hub
 height.  This particular turbine was initially designed for extreme
 cold weather applications such as the South
 Pole\footnote{https://www.nps100.com/wp/technology/\#heritage},  but other similar turbines should also be
 assessed in future analysis \cite{turbineresearch,windturbineSP}. 

\begin{figure}[htb]
	\centering
		\includegraphics[width=0.95\textwidth]{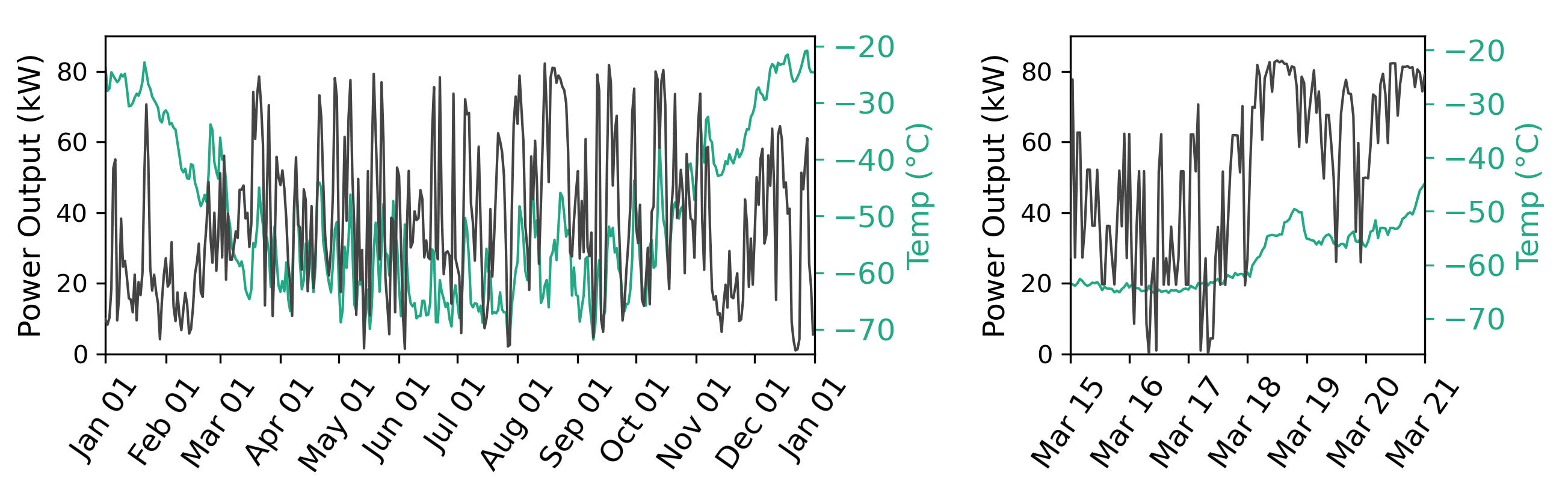}
	  \caption{(Left) Daily average of estimated power output for an NPS
            100C-24 wind turbine installed at the South Pole on a
            30-meter tower (grey).  The daily average temperature is
            shown in teal. The model uses wind speed data from 2003
            and the turbine is assumed to have a -70$^{\circ}$ C lower
            limit on operating temperature. (Right) Hourly
            average estimated power output for a six day period with
            temperature overplotted. }\label{fig5}
\end{figure}

In addition to the availability of the wind resource, an important
consideration for wind energy at the South Pole is the operational
temperature limit.  The turbine considered in this analysis operates
to -40$^{\circ}$ C, below which the turbine shuts down.  However,
initial assessments indicate turbine operation down to -70$^{\circ}$ C
is technically possible.  For the wind resource estimation, the wind
turbine is assumed to be offline when the temperature is below
-70$^{\circ}$ C, resulting in no power production during these time
periods.  The resulting expectation of power produced by a single
turbine for the conditions of the analysis year is shown in Figure
\ref{fig5}.  While this resource profile is representative, additional
atmospheric measurements at a height of at least 30 meters above
ground level would enable additional refinement and confidence. 

There are several challenges for operating turbines at
the South Pole: transportation and construction logistics, extreme
cold climate operation, and creating a suitable turbine foundation.  Turbines of the 100 kW size are typically designed to ship
within a C130 transport aircraft and to be installed with minimum
additional equipment, potentially including the use of tilt-up
towers. This is the case for the NPS 100C-24 turbine considered in
this analysis. Turbine operation in very cold climates has been
researched extensively \cite{ieawind19} and requires consideration of
materials, including metallurgy and lubricants, component heating, and
unique operating conditions in these environments. The latter includes
both specialized safety requirements and potential operational limits
during extreme weather events. 

Turbine deployment at the South Pole requires special development
for the foundation as solid ground for traditional tower footings is
not available due to the thick Antarctic ice sheet.  Development of an
ice-based foundation would be required.  Although not documented in
the literature, ice foundations have been used for smaller turbines
installed at the Summit Station in Greenland and at least twice at the
South Pole.  In all three cases a modified guyed tower foundation is
used, similar in concept to the tall meteorological towers installed
at the South Pole.  Ice-based foundations are also used for the
elevated station building \cite{foundationsnow} as well as other buildings on-site,
including the South Pole Telescope \cite{Carlstrom2011}.  In these designs the weight
of the turbine and tower is dispersed to not exceed the load bearing
weight of the ice and deadman anchors for the guy wires address the
overturning moment of the tower.   Towers of the height and weight
proposed in this study have yet to be demonstrated, however, the
engineering principles are well-understood.

\subsection{Energy Storage}
\label{subsec:es}
Energy storage (ES) is key to enabling deep decarbonization via
electrified transportation and renewable power generation.  The
evolution of contemporary energy storage is largely driven through its
role in electric vehicles (EV), which now have cost and performance
that are sufficient for mass adoption — after decades of
battery research and development.  Today, lithium-ion (Li-ion) is the
single dominant ES technology with manufacturing that is scaling
globally to meet this rapidly growing demand \cite{gesdeepdive}.
  ES is the primary strategy to address variability in solar and wind renewable resources due to its ease of implementation.

Stationary storage today is served with readily available Li-ion
batteries based on designs that were initially focused on EVs.  These serve as a
natural starting point for the evolving stationary markets but are not
optimal in several aspects.  One example is the flammability risk.
While always critical, it is of paramount importance for stationary
applications where deployment sizes are typically several orders of
magnitude larger than an EV pack and therefore can result in
significant negative outcomes.   Stationary safety today is insured
with multiple system elements to prevent such events.  Solutions to the long-term
goal of nonflammable/reduced flammability versions of Li-ion are now
emerging; one option was identified in the course of this analysis and
should remain under consideration in the future.

As the percentage of renewable generation increases in large grids, both the amount
(kWh) and type of ES needed to balance the resource variability
changes.   When the hours of available energy (duration) to ensure system
reliability surpasses 10 hours, the storage requirements enter a 
regime called long-duration energy storage
(LDES) \cite{albertus2020}.   The affordable
cost of energy storage also decreases significantly with increasing duration.  
Presently, most technologies are unable to meet LDES cost
requirements.  Li-ion is not projected to be cost effective for LDES
even as its cost continues to drop due to economies of scale for EV
mass adoption.  The LDES cost requirements create a significant
technology gap that is a great challenge for near-total renewable
generation scenarios, which attracts global attention.  For example, the United
States Department of Energy (DOE) has an earthshot
initiative to develop LDES technologies for a targeted cost reduction of 90\%
relative to Li-ion.  In the context of the analysis presented here,
Li-ion can be considered as ‘short duration’ energy storage. 
It is the dominant technology today but is not the singular
long-term solution.   A summary of the generalized tradeoffs between Li-ion and LDES is presented in Table
\ref{tbl1}.
 \begin{table}[ht]
 \caption{Summary of energy storage tradeoffs.}\label{tbl1}
\renewcommand{\arraystretch}{1.5}
 \begin{tabular*}{0.95\textwidth}{>{\raggedright}p{0.17\linewidth}
   >{\centering}p{0.23\linewidth} >{\centering}p{0.23\linewidth} >{\centering}p{0.23\linewidth}}
 \toprule
   Feature & Li-ion (short duration ES) & LDES (long duration ES) & Comment 
%\tabularnewline
 % & (short duration ES) & (long duration ES) & 
\tabularnewline
% Table header row
 \midrule
 Development stage &  Mature & Rapidly developing, heavily researched
                               %rapidly emerging 
& LDES is
                                                              a
                                                              U.S. DOE
                                                              earthshot
                                                              initiative
\tabularnewline
  Capital cost& High&High today & LDES costs are rapidly decreasing\tabularnewline
  Cycle life& Degrades with use & Generally stable & \tabularnewline
 Round trip efficiency (RTE) & Very high: $\sim$85-90\%  & Generally low: $\sim$45-60\% & \tabularnewline
Energy density& High & Unknown: generally much less than Li-ion & LDES at maturity may match Li-ion \tabularnewline
System weight& Low & Variable, often high & \tabularnewline
Flammability& High & Are typically aqueous-based and so nonflammable & Nonflammable Li-ion is emerging\tabularnewline
Low temperature stability& Good to -20$^{\circ}$C & Generally good to 0$^{\circ}$C & Many LDES are aqueous-based and will freeze\tabularnewline
 \bottomrule
 \end{tabular*}
\end{table}

The usual value proposition and design criteria derived from the energy storage cost and performance framework do not
  directly translate to the South Pole analysis presented
here, but the technology maturity and market focus directly impacts the availability
of options.  Li-ion cost and performance projections are readily
available, for example from the DOE annual report used in this
analysis \cite{esgc2022}. LDES emerging technologies are not presently
characterized with a dependable similar annual compendium.
As such, the LDES information considered here is
gathered using two approaches.  First, cost and performance values are
obtained from detailed discussions with specific LDES companies that
are in early large-scale deployments and thus can offer projections
for the next 5-10 years. Second, DOE earthshot cost targets
  are used as a guide to general aspirational values
  \cite{esgcroadmap}.  Projected values from the companies were
compared to the DOE target and in some cases found to be approximately
consistent.  The data from these companies
are therefore used for two cases of the
system-wide optimization described in the following section.

Total ES cost is a combination of up-front capital cost and the
shipping cost to the South Pole site.  The latter depends on the
energy density (in Wh/kg) of the ES technology.  There is a large
disparity today in energy density for Li-ion and LDES technologies.
Li-ion was initially developed for consumer electronics and transportation where high energy
density is critically important whereas LDES is optimized for lower
cost but not for energy density.  Li-ion generally has higher energy
density compared to LDES and therefore significantly lower total
weight and subsequent shipping cost. Additional
discussion of these cost impacts can be found in Section
\ref{subsec:assumptions}.   As a result of these considerations, we evaluate both short- and long-duration
approaches to energy storage in this analysis. 

\section{Optimization of the Hybrid System using
  REopt}\label{sec:reopt}

The REopt\texttrademark \, techno-economic decision support platform is used to
analyze options for a South Pole hybrid energy system \cite{reopt}.
REopt's development has occurred over 15 years, evolving from its
inception as spreadsheet model to a complex, open-source, mixed-integer
linear program accessible as an API or through a web-based user
interface.  REopt identifies the optimal mix of renewable energy,
conventional generation, and energy storage technologies to meet
objectives of cost savings, resilience, emissions reductions, and
energy performance. REopt generates least-cost options for serving an
identified load based on costs and performance of generation
estimations for the South Pole. The technologies considered in this
analysis include diesel power, solar photovoltaics (PV), wind turbine
generators (WTG), and both short- and long-duration electricity
battery energy storage systems (BESS).

The South Pole model has one-hour time steps for a single year of
load, wind, and solar resources. PV, WTG, and BESS technology sizes
and the hourly system dispatch are optimized to minimize total
lifecycle costs. In this analysis, a 15-year economic analysis period
is used. Costs and savings occurring in the analysis period are
discounted using standard economic principles to determine life-cycle
cost and net present value, defined as the lifetime savings compared
to an all-diesel case, for each scenario. A simple payback period is
also calculated. Finally, key outputs on avoided diesel fuel
consumption and emissions are generated.

\subsection{Overview of Scenarios}
\label{subsec:scenarios}
A cost-benefit analysis for multiple configurations of renewable technology is performed.  To create a baseline for comparison, REopt is first run assuming all power is provided by an existing central diesel plant. Several different combinations of both generation and storage technologies are explored including configurations with and without the use of diesel fuel.  Table \ref{tbl2} defines the mix of technologies for each of the scenarios modeled.  All scenarios are compared to the business-as-usual (BAU) case where power is only supplied by an existing diesel power plant. 

\begin{table}[ht]
\caption{Configuration of renewable energy generation and energy storage technology for each scenario modeled.}\label{tbl2}
\begin{tabular*}{0.5\textwidth}{@{}LCCCCCC}
\toprule
&\multicolumn{6}{C}{Scenario Label} \\ 
Technology & BAU & A & B & C & D & E  \\ % Table header row
\midrule
 Diesel &  $\bullet$ &$\bullet$ &$\bullet$ & $\bullet$& &$\bullet$ \\
 PV & &$\bullet$ & & $\bullet$&$\bullet$ &  $\bullet$\\
 WTG & & &$\bullet$ &$\bullet$ &$\bullet$ & $\bullet$\\
 Li-ion BESS && $\bullet$ &$\bullet$ &$\bullet$ & $\bullet$ &\\
 LDES BESS & & &  & & &$\bullet$\\
\bottomrule
\end{tabular*}
\end{table}

Renewable energy and BESS scenarios that include diesel generation are
also modeled with and without constraints on total diesel consumption
to inform cost-benefit considerations of achieving higher shares of
renewable energy contribution above the least-cost
solution. Additionally, sensitivity to diesel fuel cost, diesel plant
fuel economy, and LDES performance and cost assumptions are
evaluated.  These sensitivities are further discussed in Section \ref{subsec:sensitive}.

\subsection{System-wide Assumptions}
\label{subsec:assumptions}
A steady-state power demand of 170 kW is
  modeled across a full year.  This level is a representative example
  that is selected
  based on the predicted load from a future large scientific experiment
  planned for the South Pole station\footnote{private communication}. The renewable energy generation profiles for solar and wind described
in Section \ref{sec:model}  are input into the techno-economic optimization.  The
technical performance and costs of all generation options are also
entered into REopt.  Key costs and economic parameters for this
analysis are included in Table \ref{tbl3}.   Costs include the total installed
cost estimates and on-going maintenance cost estimates to assure the
systems will operate reliably after commissioning. PV and Li-ion BESS
costs are based on cost models that inform NREL’s Annual Technology
Baseline \cite{pvbenchmark2022}.  These are bottoms-up models that
contain all major system and installation components for a given
architecture. Parameters include  (but are not limited to) equipment cost, installation time and
labor rate estimates, overhead, contingency, profit, and sales tax.
Each parameter was reviewed for the South Pole specific context with
key adjustments made to labor rates and durations, development costs,
profit, and added transportation cost. WTG capital cost is separately estimated based on a NPS
100C-24 model.  
\begin{table}[ht]
 \caption{Cost estimates and system-wide assumptions used in REopt analysis.}\label{tbl3}
\renewcommand{\arraystretch}{1.5} 
\begin{tabular*}{\tblwidth}{p{0.2\linewidth}>{\centering}p{0.2\linewidth}>{\centering}p{0.17\linewidth}>{\centering}p{0.28\linewidth}}
 \toprule
   Parameter & Value & Annual Maintenance Cost & Additional Factors  \tabularnewline % Table header row
 \midrule
  Power demand& 170 kW & & Constant \tabularnewline
 Diesel fuel cost & \$40/gallon delivered & &  2.7\% annual escalation
  rate\tabularnewline
 Diesel plant fuel \newline economy & 12 kWh/gallon & & Marginal fuel economy\tabularnewline
 PV cost & \$5,330/kW-DC installed & \$42.50/kW-DC& 0.5\% annual degradation\tabularnewline
 Wind turbine cost & \$9,670/kW installed & \$230/kW & \tabularnewline
BESS, Li-ion cost & \$1,910/kW \newline+ \$840/kWh installed & & 97.5\% RTE,
                                                         DC-DC\newline 96\%
                                                         inverter \&
                                                                     rectifier efficiences
   \newline 20\% minimum state-of-charge\tabularnewline
BESS, LDES cost &\$1,810/kW\newline+ \$860/kWh installed & & 55\% RTE,
                                                            DC-DC
                                                            \newline 96\%
                                                         inverter \&
                                                                     rectifier efficiences
   \newline 10\% minimum state-of-charge\tabularnewline
Analysis period &15 years & & \tabularnewline
Discount rate& 3\% & & \tabularnewline
Inflation rate \newline& 2.5\% & &  non-fuel maintenance\tabularnewline
 \bottomrule
 \end{tabular*}
\end{table}

Costs for all renewable technologies include the
estimated procurement and total installation costs based on assumed
labor rates and hours. Labor hours for both PV and Li-ion BESS are
doubled compared to a North American standard to adjust for South Pole
conditions. Installation cost estimates for the WTG are based
  on initial turbine foundation concepts, modeled on what was used for
  the McMurdo Station wind turbines, including shipping for pre-cast
  turbine footings and extended installation timelines. As with PV and 
  Li-ion BESS, a significant cost premium is added to account for
  deployment at the South Pole including shipping, additional
  installation costs, and increased labor rates and hours. An
  additional premium is also added to standard annual operations and
  maintenance costs, including expanded equipment monitoring which
  would be required at such a location.  LDES
total installed costs are developed by consulting with industry to
estimate price at 'factory gate' and then entering those procurements
costs into the same bottoms-up model used for PV and Li-ion previously
mentioned. Although the LDES energy storage procurement prices are
estimated to be significantly lower than Li-ion battery packs, when
balance of system costs, including power electronics, engineering,
shipping to the South Pole, and installation labor are included, their
total installed costs are similar. Finally, the cost of
diesel power generation is limited to the cost of the fuel, which is
roughly estimated based on a 2012 cost \cite{brpr2012} and subsequent inflation
rates to extrapolate to 2023.  Fuel economy is based on evaluation of
vendor catalogs for diesel engine-generators and review of remote
power system performance in rural Alaskan communities\footnote{https://www.akenergyauthority.org/What-We-Do/Power-Cost-Equalization/PCE-Reports-Publications}.   

Total installed costs also include South Pole shipping costs based on
estimated weights of the equipment.  The contribution of shipping to
the South Pole to the total installed cost is significant. Table
\ref{tbl4} breaks out the shipping costs from procurement and
installation assuming \$19.8/kg for delivery to the South Pole site,
which is roughly estimated using standard inflation factors and cost
from previous economic analysis of cargo transportation \cite{traverse}. The incremental shipping costs for BESS power
electronics are assumed to be captured in the energy storage capacity
(kWh) shipping costs.
 \begin{table}[ht]
\renewcommand{\arraystretch}{1.2} 
 \caption{Breakdown of technology cost into procurement and shipping.}\label{tbl4}
 \begin{tabular*}{\tblwidth}{p{0.18\linewidth}>{\centering}p{0.15\linewidth}>{\centering}p{0.2\linewidth}>{\centering}p{0.13\linewidth}>{\centering}p{0.18\linewidth}}
 \toprule
  Technology & Procurement \& installation cost, less shipping &
                                                                 Estimated
                                                                 shipped
   weight (kg per unit) & Shipping cost & Total modeled installed cost\tabularnewline % Table header row
 \midrule
PV (\$/kW-DC) & \$2,090 & 160 & \$3,240 & \$5,330\tabularnewline
 Wind (\$/kW-AC)&  \$3,900 & 290& \$5,770 & \$9,670\tabularnewline
 BESS Li-ion (\$/kW)  & \$1,910 & 0 & \$0 & \$1,910 \tabularnewline
 BESS Li-ion (\$/kWh)  & \$680 & 8 & \$160 & \$840\tabularnewline
 BESS LDES (\$/kW) & \$1,810 & 0 & \$0 & \$1,810\tabularnewline
BESS LDES (\$/kWh) & \$370 & 25 & \$490 & \$860 \tabularnewline
 \bottomrule
 \end{tabular*}
\end{table}

For additional context of the economic environment, the levelized
costs of energy (LCOE) for diesel generated electricity, PV, and wind
power are shown in Table \ref{tbl5}. LCOE is a common cost metric to allow
comparison across technology options.  Levelized cost calculations
generally include the total installed costs, maintenance costs, annual
production, and useful life of the asset. However, the LCOE for diesel
generation shown in Table \ref{tbl5} is based on the fuel costs alone. It does
not include any incremental investment that might be needed for
additional diesel generation capacity nor non-fuel generator
operations and maintenance costs.

\begin{table}[ht]
\caption{Summary of levelized cost of energy for each generation technology.}\label{tbl5}
\begin{tabular*}{0.5\linewidth}{@{}LC@{}}
\toprule
  Technology& Levelized Cost of Energy \\
&  (\$/kWh)  \\ % Table header row
\midrule
 Diesel generation, fuel only& \$4.09\\
 PV &\$0.23 \\
 Wind & \$0.33 \\
\bottomrule
\end{tabular*}
\end{table}

The LCOEs for South Pole demonstrate how expensive electricity
currently is there due to the high costs for diesel fuel and thus the
economic impetus for integrating PV and wind power.  The economic
model shows the value of integrating BESS to facilitate greater shares
of PV and wind to offset the more expensive diesel
generation. Integrating BESS will also provide additional critical
services in a power system that this model does not consider,
including, among others, ramp rate management of variable renewable
energy, operating reserve to cover resource uncertainty, short circuit
contribution for system protection devices, and improved power
quality.

\section{Results of REopt Analysis}\label{sec:results}

\begin{figure}[htb]
	\centering
		\includegraphics[width=5.5in]{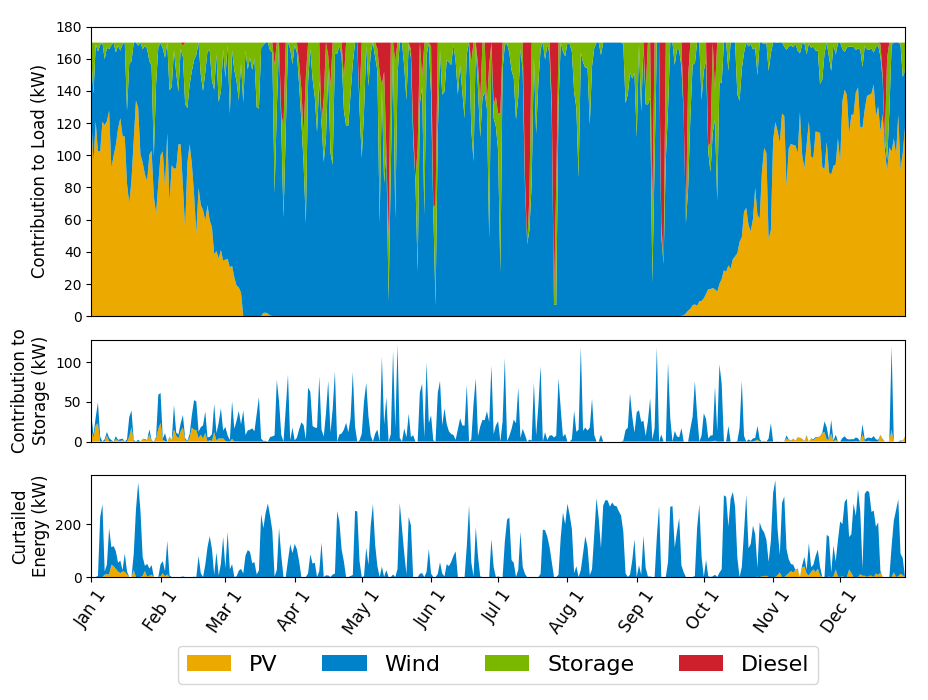}
	  \caption{(Top) Dispatch plot showing the contribution to the load over a year for scenario C with diesel, solar PV, wind, and BESS/Li-Ion. (Middle) Contribution of solar PV and wind to charging the BESS. (Bottom) Curtailed energy that could be generated based on the available resources in this scenario.}\label{fig6} 
\end{figure}

For each scenario enumerated in Section \ref{subsec:scenarios} REopt
optimizes the technology sizes needed to meet the input load over the
course of the year for the least cost.  Figure \ref{fig6} shows an
example dispatch model that meets the required load as a function of
time for scenario C.   The impact of the polar longitude is clearly
visible in the PV generated power present during the austral summer
months that then diminishes as the sun sets below the horizon.  WTG
power is available throughout the year, but BESS or diesel generated
power is required for periods of diminished wind. The bottom panel of
Figure \ref{fig6} shows the curtailed energy of the system.  This is excess
energy that could be generated in addition to that dispatched to the
electrical load or used to recharge the BESS.  For the purpose of this
analysis no economic value is given to curtailed energy. 

\begin{table}[ht]
\renewcommand{\arraystretch}{1.2} 
\caption{Summary of results from REopt optimization.}\label{tbl6}
\begin{tabular*}{\tblwidth}{@{}L >{\centering}p{0.09\linewidth}>{\centering}p{0.09\linewidth}>{\centering}p{0.09\linewidth}>{\centering}p{0.09\linewidth}>{\centering}p{0.09\linewidth}>{\centering}p{0.09\linewidth}}
\toprule
& Diesel & Diesel, PV, Li-ion & Diesel, Wind,  Li-ion & Diesel,  PV, Wind, Li-ion & PV, Wind, Li-ion & Diesel, PV, Wind, LDES\tabularnewline
 & BAU & A & B & C & D & E\tabularnewline
 % Table header row
\midrule
Life-cycle cost (\$M, discounted) & \$72.8 &\$47.5&\$18.9& \$14.9 & \$19.4 &\$15.9\tabularnewline
 Net present value (\$M) & - & \$25.3& \$53.9& \$57.8 & \$53.3 &\$56.8 \tabularnewline
 Capital expenditure (\$M) & -& \$3.8 &\$10.7& \$9.7 & \$17.4 &
                                                                \$8.9\tabularnewline
Simple payback (years) &- & 1.9 & 2.4 & 2.1 & 3.6 & 2.0\tabularnewline
PV system size (kW-DC) &-&680& -& 180 &120 & 200 \tabularnewline
Wind system size (kW) &-&-& 780 & 570 & 600 & 580\tabularnewline
BESS power (kW) &-& 50& 200& 180 &180 & 200 \tabularnewline
BESS energy (kWh) & -& 110& 3,310 & 3,410 &12,570 &2,210\tabularnewline
Hours of storage & -&2.3& 16.9 & 18.9 & 70.1 & 10.9\tabularnewline
Annual fuel consumption (gallons) & 124,000& 73,700 & 9,500 & 5,600 & 0 &8,500\tabularnewline
Fuel reduction & 0 & 41\% & 92 \% & 96\% & 100\% &93\%\tabularnewline
Annual avoided CO$_2$ (metric tons) & 0& 510& 1,170 & 1,210 & 1,270&
                                                                     1,180\tabularnewline
\bottomrule
\end{tabular*}
\end{table}

\begin{figure}[t]
	\centering
		\includegraphics[width=6.5in]{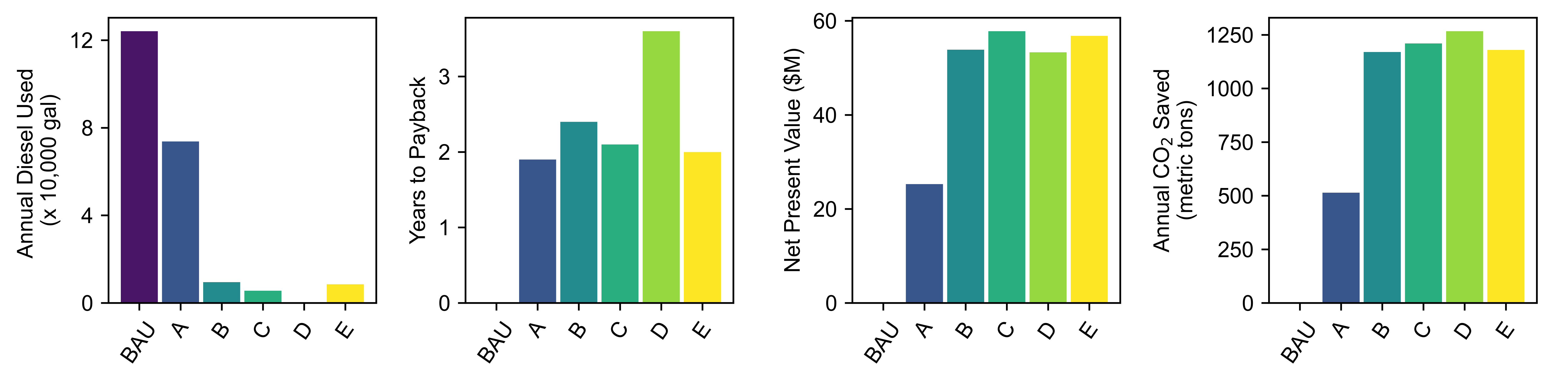}
	  \caption{Comparison of optimization results between the scenarios. }\label{fig7}
\end{figure}

Results from modeling all technology configurations are presented in Table \ref{tbl6} and Figure \ref{fig7}.  The baseline BAU configuration without integrating renewable energy and energy storage is estimated to require 124,000 gallons of fuel per year to meet the 170 kW electrical power requirement.  Over a 15-year period, the BAU fuel costs are estimated to be \$60 to \$73M in present value terms. These are the benchmark parameters against which renewable energy and energy storage scenarios are compared and are shown in the first column of Table \ref{tbl6}.  
Analysis results indicate that the all-diesel, BAU scenario, has the
highest life-cycle cost. All architecture options with renewable
energy reduce life-cycle costs below BAU and therefore result in
positive net present values. The high cost of diesel fuel is driving
high penetration of renewable energy in combination with BESS and results
in estimated payback periods of two to four years.  

Techno-economic models are a useful first step identifying technology options, indicating potential capacities, and for generating estimates of high-level costs and benefits. Techno-economic models are often revised iteratively to increase fidelity as projects move from concept, through detail design, and construction. South Pole conceptual results indicate that both PV and WTG are highly cost-effective, and inclusion of BESS should be considered to allow greater contributions of renewable energy.

The least-cost scenario shown in Table \ref{tbl6} is scenario C. Scenario C
includes 180 kW-DC of PV, a total of six 100 kW wind turbines
(rounding up the model result of 570 kW), and 180 kW / 3,400 kWh
Li-ion BESS integrated with the existing diesel system. It has a net
present value exceeding \$57M. The \$10M initial investment reduces
diesel power consumption by 96\% and has an approximately 2-year
simple payback.

Scenario A only allows PV for renewable energy generation.  Solar
radiation is at a very low level of completely unavailable for a
large fraction of the year, therefore, complete replacement of diesel
generation with only PV is not possible.  Excluding
WTG reduces net present value by 56\% below the best option, scenario
C, but it is still a highly cost-effective option with a net present
value of \$25M. The scenario A PV system is almost four times larger
than scenario C, at 683 kW-DC.   The power generated by a system of a
given size drops dramatically before sunset and after sunrise and a
larger system is required to meet the same electrical load. The LCOE
of PV is significantly lower than that of diesel and therefore when
optimizing for lowest life-cycle cost REopt determines that is less
expensive to build a larger PV system than supplying diesel during
these periods.  The BESS in this scenario is much smaller, around 50
kW for two hours, primarily serving to buffer fluctutions in solar
generation due to cloud cover. Fuel savings in this scenario are about
half as much as scenario C, 41\% savings versus 96\% respectively, but
are still significant compare to BAU.

% Figure

Similarly, scenario B only allows WTG for renewable energy
generation. A total of eight turbines are required (rounding up the
780 kW) as well as a Li-ion BESS similar in size to that of scenario
C.  As seen in the PV-only scenario A, excluding an energy generation
technology from the least-cost option reduces the net-present
value. However, a WTG-BESS diesel hybrid system is significantly
closer in total benefit to the least-cost scenario C with an NPV of
\$54M and fuel savings of 92\% because wind-generated energy is available year round.

Scenario D shows an assessment of how the optimization results change if
diesel-generated power is not allowed, i.e. the system is fully renewable
with PV, WTG and Li-Ion BESS. To eliminate diesel fuel usage, the
model slightly adjusts the PV and WTG sizing relative to the case
where diesel is allowed. However, this scenario adds significantly
larger BESS energy capacity of 12,600 kWh, more than 3.5$\times$ the
capacity of a system with diesel power (scenario C). The initial
capital investment of \$17.4M is the highest of the investment
scenarios, primarily due to the larger BESS.  The system is still
highly cost-effective with an NPV of \$53M, approximately 8\% less
than that of the system with diesel.  This can be considered the
incremental cost to increasing the reduction in fuel consumption from
96\% to 100\%, and avoiding an additional 57 metric tons of CO$_2$
emissions annually.

Finally, Scenario E captures the projected costs and benefits of using
LDES instead of Li-ion BESS.  The results across all scenarios, with
the exception of A, trend towards longer duration energy storage
motivating inclusion of LDES in this scenario.   As shown in Table
\ref{tbl4}, LDES capital cost is approximately 55\% the cost of Li-ion
before shipping (\$370 versus \$680/kWh, respectively) but is assumed
to weigh a factor of three more than Li-ion on a per kWh basis. Given
the high costs of shipping to the South Pole, LDES total installed
costs are slightly higher than Li-ion BESS. Combined with the lower
assumed round-trip energy storage efficiency (shown in Table
\ref{tbl2}) LDES has reduced economic benefit  relative to Li-ion. The cost optimal LDES solution is a smaller energy storage
system than Li-ion (2,210 kWh versus 3,410 kWh) in scenario C and has
a marginally lower NPV.  Thus, the low RTE and energy density of LDES
presently outweigh the significantly lower purchase price compared to
Li-ion.  As discussed in Section \ref{subsec:es}, LDES is an
emerging technology and ongoing evaluation of the
  techno-economic performance is warranted as it matures.

\subsection{Sensitivity to Assumptions}
\label{subsec:sensitive}

Several other scenarios have been explored with REopt to explore the
sensitivity of the results to input assumptions and constraints on the
renewable system.   Of particular importance are the assumptions on
cost of fuel at the South Pole as well as the fuel economy of diesel
power generation.  Two additional scenarios are explored, first with
cost of delivered diesel fuel reduced to \$30/gallon (scenario C1) and
second with both reduced fuel cost and increased fuel economy of 14.6 kWh/gallon for displaced diesel generation (scenario C2) \cite{mason2007}. In both scenarios, the renewable technology system includes PV, WTG, and Li-ion BESS.

% Figure
\begin{figure}[ht]
	\centering
		\includegraphics[width=6.5in]{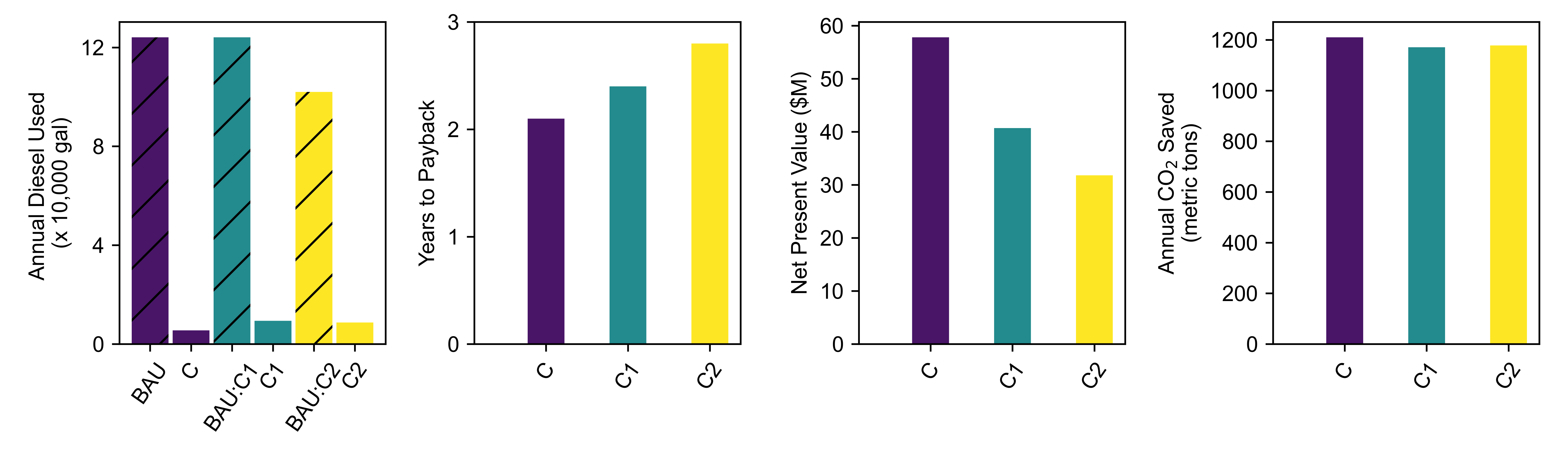}
	  \caption{Comparison to results for a hybrid renewable system with PV, WTG, Li-ion BESS (scenario C) in combination with diesel generation under different input assumptions on fuel cost and economy.  Scenario C1 corresponds to a lower fuel cost and scenario C2 includes both lower fuel cost and increased fuel economy.}\label{fig8}
\end{figure}

With lower cost fuel and more efficient diesel generation, there is still strong motivation for integrating renewable generation at the South Pole. Results for these alternative scenarios are given in Table \ref{tbl7} and a comparison shown in Figure \ref{fig8}. In scenario C1 and C2, the sizes of the PV and wind plants do not change significantly. Instead, the Li-ion BESS energy capacity is reduced by approximately a factor of two. Simple payback times remain less than three years and the annual fuel savings exceeds 90\%. While the NPV does decrease with more conservative fuel assumptions as expected, the savings is still significant at $>$\$30M. 

\begin{table}[ht]
\caption{Comparison of REopt results for variations in fuel cost and economy.}\label{tbl7}
\renewcommand{\arraystretch}{1.2} 
\begin{tabular*}{0.7\textwidth}{L >{\centering}p{0.09\linewidth} >{\centering}p{0.09\linewidth} >{\centering}p{0.09\linewidth}}
\toprule
  & \multicolumn{3}{C}{Diesel, PV, Wind, Li-ion BESS} \tabularnewline % Table header row
& C & C1 & C2\tabularnewline
\midrule
 Net present value (\$M) & \$57.8 & \$40.7 & \$31.8 \tabularnewline
 Capital expenditure (\$M)& \$9.7 & \$7.9 & \$7.5 \tabularnewline
Simple payback (years) & 2.1 & 2.4 & 2.8\tabularnewline
PV system size (kW-DC) & 180 & 190 & 200\tabularnewline
Wind system size (kW) & 570 & 520 & 500 \tabularnewline
BESS power (kW)  & 180 & 160 & 140\tabularnewline
BESS energy (kWh) & 3,410 & 1,820 & 1,570 \tabularnewline
Hours of storage & 18.9 & 11.8 & 11.4 \tabularnewline
Annual fuel consumption (gallons) & 5,600 & 9,400 & 8,700\tabularnewline
Fuel reduction &  96\% & 92\% & 92\%\tabularnewline
Annual avoided CO$_2$ (metric tons) & 1,210 & 1,170 & 1,180\tabularnewline

\bottomrule
\end{tabular*}
\end{table}

Another sensitivity explored is the sizing of the PV system, which is
nearly a factor of four larger in scenario A compared to scenario C
driven by periods of low solar resource availability.  To this end, an
artificial constraint is imposed to only include the austral summer
months between November 1 to Jan 31, when the sun is farther above the
horizon.   A hybrid system with PV and diesel energy generation and
Li-ion BESS is modeled under this constraint, resulting in a PV system
size of 350 kW and a 30 kWh BESS.   A second model is then generated
with the technology sizes fixed at these values but calculating the
energy generated over the full year. This smaller system will reduce
diesel consumption by 36\% (compared to 41\% in scenario A), with a
capital cost of \$1.9M and an NPV of \$23.8M.    This result
indicates that a majority of the benefit from a hybrid system with PV
only renewable generation can be reaped from a smaller system with a
lower capital cost. 

Finally, the sensitivity to the cost of LDES is considered by
  modeling a futuristic case where the LDES energy cost is
  reduced by $\sim$60\% compared to scenario E.  The energy density is
  also assumed to be a factor of nine times denser.   As noted in
  Section \ref{subsec:assumptions}, the total installed cost for LDES
  is dominated by balance of system and shipping costs.  Under these
  assumptions, the optimized model is a system with PV and WTG sizes
  similar to scenario E.  The size of the LDES system increases
  significantly, both in power and energy capacity, although somewhat
  surprisingly reducing the diesel consumption only by an additional
  2.6\% and slightly increasing NPV.  This scenario is economically
  very similar to the more conservative LDES example suggesting that
  the system optimization is dominated overall by the energy
  generation components. 

\section{Conclusions}\label{sec:conclusions}
In this work we present a feasibility analysis for renewable power
generation at the South Pole.    Detailed solar and wind resource
profiles for one year are generated using on-site meteorological data.
A techno-economic optimization is performed using these profiles as
well as highly tailored economic inputs, modeling the least-cost
solution to generate 170 kW of electrical power for a period of 15
years.  A hybrid renewable system consisting of PV solar panels, wind
turbine generators, a Li-ion energy storage system integrated with an
existing diesel system is able to reduce diesel fuel consumption by
95\% resulting in a net present value of \$57M.   Such a system would
require an initial capital investment of \$9.7M dollars with a simple
payback period of 2.1 years.   Several additional scenarios are
modeled, with results ranging from 40\% -100\% diesel fuel reduction
and the associated decarbonization. In all scenarios, renewable energy
generation is highly cost effective.

Realizing these benefits requires addressing several key technical
challenges in the on-site implementation of renewable energy
technology in the South Pole environment.   Future work will focus on relevant
engineering developments for each component.   The PV system modeled
here assumes the panels are installed as a standalone array as opposed
to mounted on a building.  Windblown snow creates drifts around
buildings and other structures mounted on the snow at the South Pole \cite{spsnowdrift}.   A strategy for snow drift prevention and maintenance will be
required to keep the panels fully operational and to reduce the
general impact of drifts at the site.   One of the main complications
of Li-ion energy storage is its flammable nature.  While not a unique
concern for the South Pole, fire has amplified consequences at such a
remote site.   During the course of this study, emerging non-flammable
Li-ion storage technology was identified and going forward should be
preferentially evaluated to meet the storage needs.  Additionally,
continued tracking of already identified as well as nascent LDES technologies should occur.

Several elements of wind turbine implementation require development.
As noted in Section \ref{subsec:wind} the predicted power generation
profile for a wind turbine can be improved through collection of
additional wind data at a height of 30 meters above the ground.  A
more complete wind resource assessment will also include combining
this new data with an investigation of long-term trends in the
existing meteorological data.  Wind turbines also have the potential
to produce electromagnetic interference (EMI) \cite{winkel2019}.  The South Pole Station
includes the radio-quiet Dark Sector, which hosts scientific
experiments that are EMI sensitive \cite{asmano5}.  A detailed assessment of EMI
from a turbine must be performed and a subsequent mitigation plan
including location on site or turbine modifications will be necessary.
Additionally, a fundamental assumption of the wind resource modeling
is operation down to a lower temperature limit of -70$^{\circ}$ C.
The turbine identified here was originally designed to meet that goal
but is only currently guaranteed to a temperature of -40$^{\circ}$C.
The turbine will need to be evaluated for modification and operation
at the necessary lower temperatures demonstrated.   Finally, the
foundation design for the turbine will require development to
determine the viability of a compacted snow foundation.   A devoted
campaign of engineering development has the potential to address the
technical challenges noted here, which would pave the way for
renewable energy resources and energy storage to become an integral
part of decarbonized power generation at the South Pole.  

\section*{Acknowledgements} 
Work at Argonne National Laboratory was supported under the
U.S. Department of Energy contract DE-AC02-06CH11357.  This work was
authored in part by the National Renewable Energy Laboratory, operated
by Alliance for Sustainable Energy, LLC, for the U.S. Department of
Energy (DOE) under Contract No. DE-AC36-08GO28308.  Data used
  in the
  solar modeling was obtained from the National Aeronautics and Space
  Administration (NASA) Langley Research Center (LaRC) Prediction of
  Worldwide Energy Resource (POWER) Project funded through the NASA
  Earth Science/Applied Science Program. The data was obtained from
  the POWER Project's Hourly 2.3.2 version on 2022/11/04.  We also
acknowledge use of meteorological and radiation data collected by the NOAA
Global Monitoring Laboratory (GML) at the South Pole observatory (SPO)
covering the period from 2016-01-01 to 2016-12-31  and downloaded on
2022-12-02 from the World Radiation Monitoring Center - Baseline
Surface Radiation Network (WRMC-BSRN) archive.  We also
  acknowledge the use of wind data collected by the NOAA GML at the
  SPO covering the periods including 2003-01-01 to 2003-12-30 and
  2008-02-01 to 2009-06-01.
The views expressed in the article do not necessarily represent the views of the DOE or the U.S. Government. The U.S. Government retains and the publisher, by accepting the article for publication, acknowledges that the U.S. Government retains a nonexclusive, paid-up, irrevocable, worldwide license to publish or reproduce the published form of this work, or allow others to do so, for U.S. Government purposes.

\bibliographystyle{model1-num-names}
%%below

%\bibliography{sp_feas_refs}
\bibliography{main}

\end{document}